%% file: PaperForReview.tex
\documentclass[10pt,twocolumn,letterpaper]{article}

\usepackage{cvpr}              % To produce the CAMERA-READY version
\usepackage{graphicx}
\usepackage{amsmath}
\usepackage{amssymb}
\usepackage{booktabs}
\usepackage{multirow}
\usepackage{arydshln}
\usepackage{color}
\usepackage{bm}

\usepackage[pagebackref,breaklinks,colorlinks]{hyperref}

\usepackage[capitalize]{cleveref}
\crefname{section}{Sec.}{Secs.}
\Crefname{section}{Section}{Sections}
\Crefname{table}{Table}{Tables}
\crefname{table}{Tab.}{Tabs.}

\def\cvprPaperID{28} % *** Enter the CVPR Paper ID here
\def\confName{CVPR}
\def\confYear{2022}

\begin{document}

\title{A Closer Look at Blind Super-Resolution: Degradation Models, Baselines, and Performance Upper Bounds}

\author{Wenlong Zhang{\textsuperscript{1}}, Guangyuan Shi{\textsuperscript{1}}, Yihao Liu{\textsuperscript{2 3}}, Chao Dong{\textsuperscript{2 4}}, Xiao-Ming Wu{\textsuperscript{1}}\thanks{Corresponding author}\\
{\textsuperscript{1}}The HongKong Polytechnic University,
{\textsuperscript{2}}Shenzhen Institute of Advanced Technology, CAS \\
\textsuperscript{3}University of Chinese Academy of Sciences,
\textsuperscript{4}Shanghai AI Laboratory\\
% {\textsuperscript{3}}University of Chinese Academy of Sciences\\
{\tt\small \{wenlong.zhang, guang-yuan.shi\}@connect.polyu.hk,
\{yh.liu4, chao.dong\}@siat.ac.cn} \\
{\tt\small xiao-ming.wu@polyu.edu.hk}
% \and
}
\maketitle

%%%%%%%%% ABSTRACT

%%%%%%%%% BODY TEXT

\input{latex/Abstract}

\input{latex/Introduction}
\input{latex/RelatedWork}
\input{latex/Method}

\input{latex/Analysis}

\input{latex/Experiments}

\input{latex/Conclusion}

%-------------------------------------------------------------------------

%%%%%%%%% REFERENCES
{\small
\bibliographystyle{ieee_fullname}
\bibliography{PaperForReview}
}

\end{document}

%% file: latex/Abstract.tex
\begin{abstract}

%Blind super-resolution (SR) has demonstrated the potential to handle low-quality images from the real world. 

Degradation models play an important role in Blind super-resolution (SR). The classical degradation model, which mainly involves blur degradation, is too simple to simulate real-world scenarios. The recently proposed practical degradation model includes a full spectrum of degradation types, but only considers complex cases that use all degradation types in the degradation process, while ignoring many important corner cases that are common in the real world. To address this problem, we propose a unified gated degradation model to generate a broad set of degradation cases using a random gate controller. Based on the gated degradation model, we propose simple baseline networks that can effectively handle non-blind, classical, practical degradation cases as well as many other corner cases. To fairly evaluate the performance of our baseline networks against state-of-the-art methods and understand their limits, we introduce the performance upper bound of an SR network for every degradation type. Our empirical analysis shows that with the unified gated degradation model, the proposed baselines can achieve much better performance than existing methods in quantitative and qualitative results, which are close to the performance upper bounds.

\end{abstract}

%% file: latex/Introduction.tex
\section{Introduction}
\label{section:inro}

Traditional image super-resolution (SR) aims at reconstructing a high-resolution (HR) image from a low-resolution (LR) observation. In the past decade, convolutional neural networks (CNNs) \cite{dong2014learning,zhang2018residual,zhang2018rcan,wang2018recovering} have demonstrated superior performance in this task due to their powerful representation learning ability. Unlike traditional image SR, blind SR aims to generate an HR image from the counterpart one with a variety of unknown  degradation types. 

\begin{figure}[!t]

\setlength{\abovecaptionskip}{-0.2pt}
\setlength{\belowcaptionskip}{-18pt}

  \centering
   \includegraphics[width=0.9\linewidth]{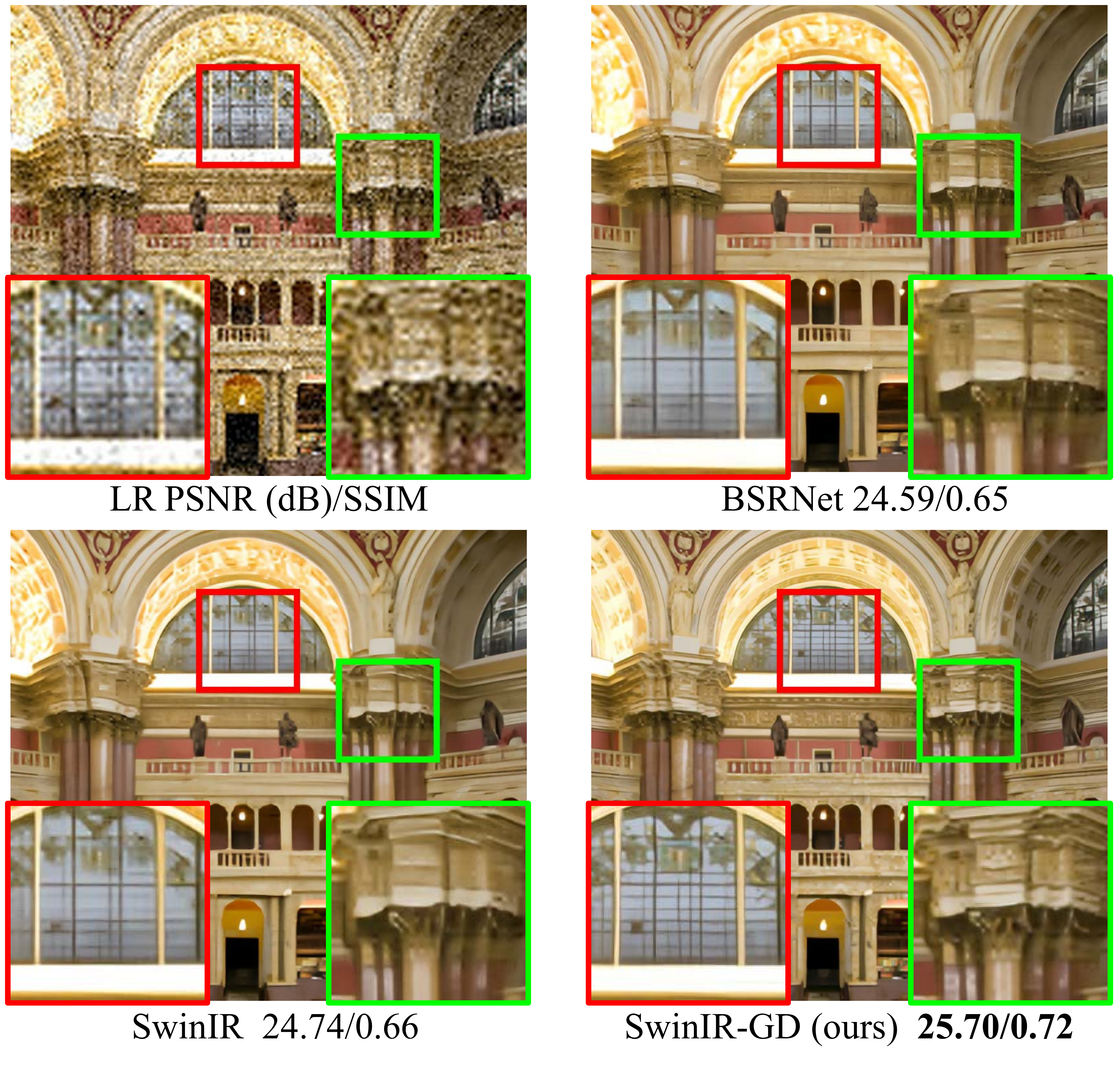}

   \caption{Visual comparisons of our method and state-of-the-art methods in $\times$4 blind  super-resolution.}
   \label{fig: intro}
\end{figure}

% Research Questions (RQ) and Challenges. The study of blind SR around two basic research questions: 1) How to learn an   (e.g., isotropic Gaussian blur)

Recent blind SR methods can be roughly divided into two groups. The first one \cite{zhang2018learning} adopts a classical degradation model, which adds a blur degradation to the non-blind degradation model. Extensive research has achieved significant progress, such as kernel estimation \cite{gu2019blind,luo2020unfolding,luo2021end}, representation learning \cite{wang2021unsupervised}, zero-shot learning\cite{shocher2018zero}, meta-learning\cite{soh2020meta,park2020fast}, optimization method\cite{cornillere2019blind}, real-world dataset\cite{cai2019toward,wei2020aim} and unsupervised methods\cite{yuan2018unsupervised,lugmayr2019unsupervised}. 
%These methods have achieved great improvement on the classical blind SR. 

% In addition, FAIG \cite{xie2021finding} finds that a blind SR network can achieve comparable performance with the state-of-the-art (SOTA) methods. 

% Gu et al. \cite{gu2019blind} propose an iterative correction algorithm with a condition network which modulates the SR output to the corresponding blur condition. Luo et al. \cite{luo2020unfolding,luo2021end} further improve the iterative correction algorithm to an end-to-end framework. Wang et al. \cite{wang2021unsupervised} proposed a representative learning method to generate a unsupervised kernel representation.    

However, down-sampling with blur degradation is still an overly simple simulation since there exist many other degradation types in the real world. To address this problem, recent research introduces a practical degradation (PD) model \cite{zhang2021designing,wang2021real} to mimic the degradation process from HR to LR images with various degradation types, including multiple blur types, noise types, and JPEG compression. 
Furthermore, BSRGAN\cite{zhang2021designing} introduces a shuffle operation to expand the degradation space, and RealESRGAN\cite{wang2021real} designs a high-order pipeline to simulate complex degradations.
% It also leads to a separation from the practical to the classical and non-blind degradation models.   on complicated degradation cases, such as the combination of blur, noise and JPEG.   (e.g., Gaussian noise, Poisson noise, camera sensor noise, etc.)   (e.g., generalized/ plateau-shaped Gaussian blur)

Despite recent progress, blind SR remains a challenging problem. In our pilot study, we have identified three key issues not well examined in previous research: \textit{1) the design of a general degradation model that can cover most or even all degradation cases; 2) strong baselines that can well handle most degradation cases; 3) the study of performance upper bounds that can be used to evaluate the performance of existing blind SR methods w.r.t. distinct degradation cases.}
\textit{\textbf{For issue 1}}, it is a well-known fact that the degradation process of real-world images is highly random, which may involve a broad set of degradation cases.
%\textit{\textbf{it is impossible that every image in the real-world contains the most complicated degradation type.}}
However, existing degradation models only cover limited degradation cases. The classical degradation model~\cite{luo2021end,wang2021unsupervised} only focuses on the blur degradation type, whereas the practical degradation model~\cite{zhang2021designing,wang2021real} considers the most complex degradation cases and ignores many other corner cases (e.g., combinations of a subset of degradation types). This leads to \textit{\textbf{issue 2}}. Due to the lack of a unified degradation model, existing methods can not perform well in various degradation cases, as shown in Figure \ref{fig: intro}. Hence, a strong baseline that can well handle different degradation cases is in need, which can facilitate the comparative analysis of the learning ability of a blind SR network.
%Without the baseline, it is difficult to determine that one or more networks are required in a specific scenario.
\textit{\textbf{For issue 3}}, there lacks the study of quantitative performance upper bounds that an SR network trained with a specific degradation type (e.g., blur 2.0) can achieve on the test dataset. Without comparison with the upper bounds, it is difficult to evaluate whether a blind SR network is good enough in a special degradation case. 
%Further, existing practical blind SR methods do not show detailed performance comparison on internal degradation cases.  
 %com the performance   both classical and practical blind SR lack quantitative performance comparisons between a blind SR network and the upper bound. Specifically, the upper bound represents the performance of a SR network with a special degradation type (e.g., blur 2.0) on the corresponding test dataset (e.g., blur 2.0). Without the upper bound comparisons, it is hard to evaluate whether the blind SR network is good enough on a special degradation. Even more, the existing practical blind SR methods do not show detailed performance comparisons on internal degradation cases.  

% Therefore, existing practical blind SR methods can not achieve good performance on most real-world internal degradation cases, due to the lack of a unified degradation model.

In this paper, we take a closer look at the three issues and provide simple yet effective solutions. \textit{\textbf{To address issue 1}}, we propose a unified gated practical degradation (GD) model for blind SR. Specifically, the proposed GD model introduces a gate mechanism that can generate various combinations of degradation types to cover 
%the space of 
as many degradation cases as possible in the real-world. In the degradation process, we use a random gate controller to determine whether the HR image undergoes a certain degradation. As such, the proposed GD model can include traditional cases (non-blind SR), simple degradation cases (classical blind SR), complex degradation cases (practical blind SR), as well as many other common corner cases. 
%In addition, some cases ignored by the practical degradation model are also included, such as single noise and JPEG. 
The GD model leads to  \textit{\textbf{solutions to issue 2}}. Based on the GD model, we propose strong baseline networks that can well handle most degradation cases. Without additional design, our blind SR networks can surprisingly achieve consistent and significant  performance gains over existing methods. \textit{\textbf{To address issue 3}}, we introduce performance upper bounds to effectively evaluate existing methods and our proposed baselines on various degradation cases. Specifically, the performance upper bound for a certain degradation case can be obtained by training an SR network on the corresponding dataset. With the performance upper bounds, we provide a comprehensive comparative analysis of a blind SR network on the classical and practical degradation models as well as our proposed GD model (section \ref{sec:analysis}). 
% The quantitative comparisons show that a baseline network with the proposed GD model can achieve performance close to the upper bounds and much better than the practical degradation model on most degradation types. 
The contributions of this paper are summarized as follows.  
\begin{itemize}
\item We propose a unified gated degradation model that can effectively handle non-blind, classical, practical degradation cases as well as many other corner cases.
\item To the best of our knowledge, we are the first to provide a comprehensive analysis of blind SR with performance upper bounds on both the classical and practical blind SR paradigms.
\item We show that the baseline networks with the proposed GD model can achieve superior performance close to the upper bounds. 
\end{itemize}

%% file: latex/RelatedWork.tex
\section{Related work}
\textbf{Non-blind super-resolution.} Since Dong et al. \cite{dong2014learning} first introduced convolutional neural networks (CNNs) to the SR task, a series of learning-based works \cite{tong2017image,zhang2012single,haris2018deep,kim2016accurate, He_2019_CVPR, kim2016accurate, lim2017enhanced, zhang2018residual,zhang2018rcan} have achieved great performance. 
% For example, residual and dense block \cite{zhang2018residual} and channel attention \cite{zhang2018rcan} are introduced to SR problem and achieve SOTA results. 
To reconstruct realistic textures, generative adversarial networks (GAN) \cite{ledig2017photo} are introduced to generate visually pleasing results. A series of GAN-based methods \cite{sajjadi2017enhancenet,wang2018recovering,Wang_2018_ECCV_Workshops,Rad_2019_ICCV,yuan2018unsupervised} are proposed to improve the visual results and quantitative results\cite{zhang2019ranksrgan,wenlong2021ranksrgan}.  
However, these methods focus on the bicubic down-sampling degradation model, which is too idealistic compared with the LR image in the real-world. 

\textbf{Classical blind super-resolution.} To enhance the reconstruction ability of the SR network in the real-world. Zhang et al. \cite{zhang2018learning} proposed a classical blind degradation model consisting of Gaussian blur and noise with a range. Furthermore, Gu et al.\cite{gu2019blind} proposed a kernel estimation method with an iterative correction algorithm. Then, DAN \cite{luo2020unfolding,luo2021end} and DASR \cite{wang2021unsupervised} are proposed to further improve the blind SR results. In addition, a series of methods achieved great improvements in classical blind SR, such as zero-shot learning\cite{shocher2018zero}, meta-learning\cite{soh2020meta,park2020fast}, optimization method\cite{cornillere2019blind}, real-world dataset\cite{cai2019toward,wei2020aim}, and unsupervised methods\cite{yuan2018unsupervised,lugmayr2019unsupervised}. However, these methods only consider a part of degradation types in the real-world. The LR images in the real-world are affected by a variety of degradation types. 

\textbf{Practical blind super-resolution.} Considering that there are multiple degradation types in the real-world. Zhang et al.\cite{zhang2021designing} proposed a practical degradation model, which includes multiple blur types, down-sampling operation (bilinear and bicubic) with a scale factor, camera noise, and JPEG compression. The degradation order is not fixed but randomly shuffled. Furthermore, RealESRGAN\cite{wang2021real} introduced a high-order operation to enhance the practical degradation model. However, these methods can achieve promising results on complex degradation while ignoring some easy cases.   

%% file: latex/Method.tex
\begin{figure*}[!t]

  \centering
   \includegraphics[width=1\linewidth]{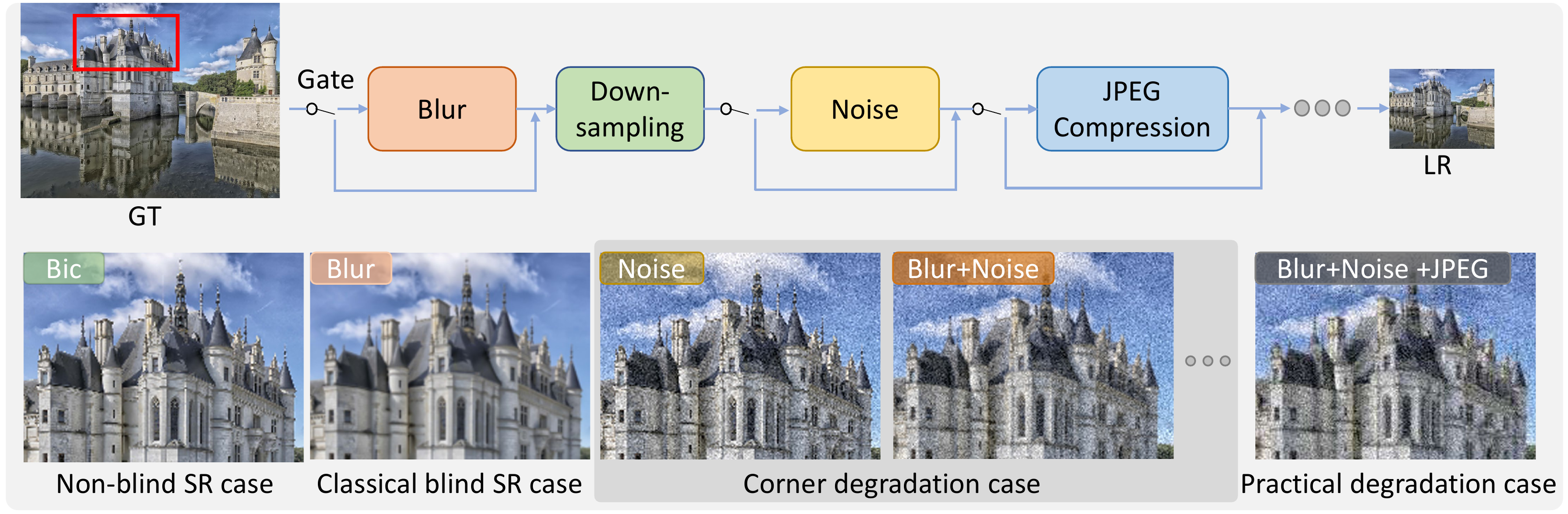}

   \caption{Our proposed gated degradation model is a unified model that encompasses non-blind SR, classical blind SR, and practical blind SR. The gate controller can generate various corner degradation cases and complex degradation cases to 
   %match the different cases in the 
   simulate
   real-world scenarios. }
   \label{fig:method}
\end{figure*}

\section{Degradation Models} % p3-p4
%\subsection{Preliminary}
\subsection{Prior Research}
\textbf{Classical Degradation Model.}
Blind SR is an ill-posed inverse problem which assumes the HR image is affected by multiple degradation types. Mathematically, the LR image $I^{LR}$ is generated from the HR image $I^{HR}$ 
%by a degradation model with multiple degradation types
as follows:
\begin{equation}
    I^{LR}= D_{k,n,j}(I^{HR}) =[(k \otimes I^{HR}) {\downarrow}_s + n]_{j}, \label{eq:4.1.1}
\end{equation}
where $\otimes$ represents convolution. First, the high-resolution image $I^{HR}$ is convolved with Gaussian blur kernel $k$. Then, the blurred image is down-sampled (denoted by ${\downarrow}_s$) and an additive white Gaussian noise (denoted by $n$) is added to the degraded image. Finally, the low-resolution image $I^{LR}$ is obtained by JPEG compression (denoted by $j$).

% Recent blind SR methods \cite{gu2019blind,luo2020unfolding, wang2021unsupervised,xie2021finding} adopt classical degradation model. Specifically, the blur kernels $k$ usually use isotropic and anisotropic Gaussian kernels with a range (e.g 0.2-4.0). Down-sampling operation ${\downarrow}_s$ is bicubic. 
With the classical degradation model, existing blind SR methods~\cite{gu2019blind,luo2020unfolding, wang2021unsupervised,xie2021finding} focus on the blur degradation while using a fixed noise (e.g., $n=20$) rather than a range noise (e.g., $n\in[1, 30]$). JPEG compression is generally not considered. Note that without the blur, noise, and JPEG degradation, the classical degradation model is equivalent to the non-blind degradation model.

\textbf{Practical Degradation Model.}
Different from the classical degradation model, the practical degradation model \cite{zhang2021designing,wang2021real} assumes the HR image undergoes a series of degradation cases to generate the LR image:
\begin{equation}
    I^{LR}= D_{p}(I^{HR}) =(D_1 \circ D_2 \circ D_3 \cdots D_m) ( I^{HR}),
\end{equation}
% \begin{equation}
% \begin{aligned}
%     &I^{LR}= D_{p}(I^{HR}) =(D_1 \circ D_2 \circ D_3 \cdots D_i) ( I^{HR}) \\
%     &where, D_i \in \{D_k, D_n, \dots, D_{n,j}, \dots, D_{k,n,j}, \dots\},\\
% \end{aligned}
% \label{eq:4.1.1}
% \end{equation}
where $D_p$ denotes the practical degradation process and $ D_i \in \{D_k, D_n, D_{j}, \cdots\}$, $\forall i\in \{1,\ldots, m\}$, represents a base degradation type,
%that sub-degradation model from the classical degradation model,
e.g., $D_k$ is the blur degradation, and $D_{n}$ is the noise degradation.

To simulate more complex degradation cases, the degradation models in BSRGAN~\cite{zhang2021designing} and RealESRGAN~\cite{wang2021real} use a wide range of base degradation types including multiple blur types (e.g., generalized Gaussian blur and plateau-shaped Gaussian blur), multiple down-sampling schemes (e.g., nearest, bilinear, and bicubic), and multiple noises (e.g., Poisson noise and camera sensor noise).
%These designs can further enhance the degradation model to cover the degradation space in the real-world.  

% Specifically, BSRGAN adopts $D_{practical}$ with a shuffle operation to generate LR image and RealESRGAN introduces a high order degradation process. These designs can further enhance the degradation model to cover the whole degradation space in the real-world. 

\subsection{Our Proposed Gated Degradation Model}
% The practical degradation model achieves considerable progress in handling degradation in the real-world. However, it fails to handle internal cases, i.e., combinations of  such as single blur, noise, or other internal degradation types. In other words, the practical model focuses on complicated degradation while ignoring the easy internal cases. 

The practical degradation model only considers complex degradation cases by using \emph{all} (or most) base degradation types in the degradation process. However, it
%\emph{full} combination of all degradation types (every single degradation type must be used in the degradation process) 
ignores important \emph{corner} cases, i.e., combinations of different subsets of base degradation types, which are prevalent in the real world. 
%There lacks the sub-degradation models and the sub-combination of all sub-degradation models in the whole degradation space.
Motivated by this, we propose a unified degradation model by introducing a gate mechanism to \emph{randomly select} the base degradation types to be included in the degradation process. Formally,
% \begin{equation}
% \begin{aligned}
%     I^{LR} &= D_{g}(I^{HR}, gate) \\
%     &= (g(D_1) \circ g(D_2) \circ g(D_3) \cdots g(D_i)) (I^{HR}) \\
%     &\! \! \! \! \! \! \! \!  \! \! \! \! where, D_i \in \{D_k, D_n, \dots, D_{n,j}, \dots, D_{b,n,j}, \dots\},\\
% &\! \! \! \! \! \! \! \!  \! \! \! \! \left\{    
%     \begin{aligned}
%     &g(D_i)(I^{HR}) = D_i(I^{HR}), \qquad   gate =1 \\
%     &g(D_i)(I^{HR}) = I^{HR}, \qquad \qquad gate =0
%     \end{aligned}
% \right.
% \end{aligned}
% \label{eq:4.1.2}
% \end{equation}
\begin{equation}
    \begin{aligned}
        I^{LR} &= D_{g}(I^{HR}) \\
        &= (\sigma_{g}(D_1) \circ \sigma_{g}(D_2) \circ \sigma_{g}(D_3) \cdots \sigma_{g}(D_m)) (I^{HR}), \\
    \end{aligned}
\end{equation}
where $D_{g}$ denotes the gated degradation process, and $D_i \in \{D_k, D_n, D_j, \cdots\}$, $\forall i\in \{1,\ldots, m\}$, represents a base degradation type. The gate controller $\sigma_{g}$ determines whether $D_i$ is used in the degradation process, i.e.,
\begin{equation}
    \sigma_{g}(D_i)(I^d) = \begin{cases}
     D_i(I^d) ,& g =1, \\
     I^d ,& g =0,
    \end{cases}
\end{equation}
where $I^d$ denotes the degraded (or input) HR image. Note that when all gates $g=1$, the gated degradation model is equivalent to the practical degradation model, whereas when all the gates $g=0$, it is the same as the traditional non-blind SR. The gate controller allows to generate various combinations of base degradation types, and hence our degradation model is a unifed model that encompasses non-blind SR, classical blind SR, and practical blind SR.

%% file: latex/Analysis.tex
\section{A Comprehensive Analysis of Blind SR with Performance Upper Bounds} % p4-p5.5
\label{sec:analysis}

This section analyzes blind SR networks with the existing classical, practical, and proposed gated degradation model. We find that a blind SR network can achieve promising performance with our proposed gated degradation model, while the blind SR network with a practical model has a significant performance drop.  

%\subsection{Preliminary, Upper Bound and Settings}
% Recently blind super-resolution is divided into two groups. The first group IKC\cite{gu2019blind}, DAN\cite{luo2020unfolding,luo2021end},and DASR\cite{wang2021unsupervised} are based on classical degradation model, such as isotropic blur, anisotropic and blur with a single fixed noise. The second group Real-ESRGAN\cite{wang2021real} and BSRGAN\cite{zhang2021designing} tend to construct a complex real-world degradation model that includes multiple types of blur, noise, compression, and high-order degradation.

\textbf{Preliminary.} FAIG\cite{xie2021finding} shows a one-branch blind SR network can achieve comparable results compared with the SOTA methods DAN\cite{luo2020unfolding} and DASR\cite{wang2021unsupervised}. In the BSD100 \cite{martin2001database} validation dataset with blur degradation type, the performance of a one-branch network is higher than SOTA DAN and DASR, about 0.05 dB. So, the one-branch network is considered as a base network to analyze the blind SR problem. Similarly, RealESRGAN\cite{wang2021real} and BSRGAN\cite{zhang2021designing} adopt a powerful one-branch network RRDBNet as the blind SR network to solve the practical blind SR problem.

\textbf{Performance Upper Bound.} An essential issue of the practical blind SR is how to evaluate blind SR networks effectively. Based on the proposed gated degradation model, the performance upper bound can easily be introduced to clearly evaluate the blind SR network. Take a special degradation type bicubic as an example. To get the upper bound, we could train a special SR network with bicubic type and test the well-trained network on the corresponding bicubic test dataset. A similar procedure can obtain the upper bounds of other corner degradation types. The definition of the upper bound is a vital tool to evaluate blind SR.

\textbf{Setting.} In this section, RRDBNet is used as the primary blind SR network (BSRNet), which is trained on a representative degradation model. The degradation model includes isotropic Gaussian blur [0.1, 3.0], additive Gaussian noise [1, 30], and JPEG [40, 95]. To clearly evaluate the BSRNet, we design a validation dataset \textit{Practical8}, which includes every corner degradation case - \{bic, b2.0, n20, j60, b2.0n20, b2.0j60, n20j60, b2.0n20j60\}. Then, we train 8 SR models to get the upper bound on every corner case. Therefore, we can use the PSNR distance between BSRNet and upper bound for the evaluation on \textit{Practical8}. We adopt a similar setting for the classical degradation model.         

% a practical degradation model (PD) and the proposed gated degradation model are employed to evaluate the blind SR network quantitatively. RRDBNet\cite{wang2021real,zhang2021designing}, which is a backbone in SOTA practical blind SR, is used as the base blind SR network (BSRNet). Other experimental settings are consistent.
% The BSRNet-PD is trained with a light practical degradation model which includes isotropic Gaussian blur [0.1, 3.0], noise [1, 30] and JPEG [40,95]. The BSRNet-GD is trained on the same setting as the proposed gated degradation model. The upper bound is obtained by the corresponding corner degradation model based on equation \ref{eq: UpperBound}.

% (more specifications are in the supp. material).    

\subsection{Analysis of Classical Blind SR}
Similar to FAIG \cite{xie2021finding}, we train the BSRNet-FAIG on a classical degradation model with isotropic Gaussian blur [0, 3.0]. Then, we train 5 SR networks to get the corresponding upper bound on the validation dataset with bicubic (bic) and blur $\{$0.6, 1.2, 1.8, 2.4$\}$.

\input{latex/tabel_analysis}

\vskip -0.25cm

From Table \ref{tab: classicalblindSR}, we find that BSRNet has a slight performance drop (about 0.3 dB) on PSNR compared with the corresponding upper bound. The slight performance drop is relatively acceptable on the blind SR problem since it is more challenging than the non-blind SR. This exciting observation motivates us to investigate the underlying learning ability of blind SR networks, especially on a practical degradation model.

\begin{figure}[h]
\setlength{\belowcaptionskip}{-0.35cm}

  \centering
   \includegraphics[width=1\linewidth]{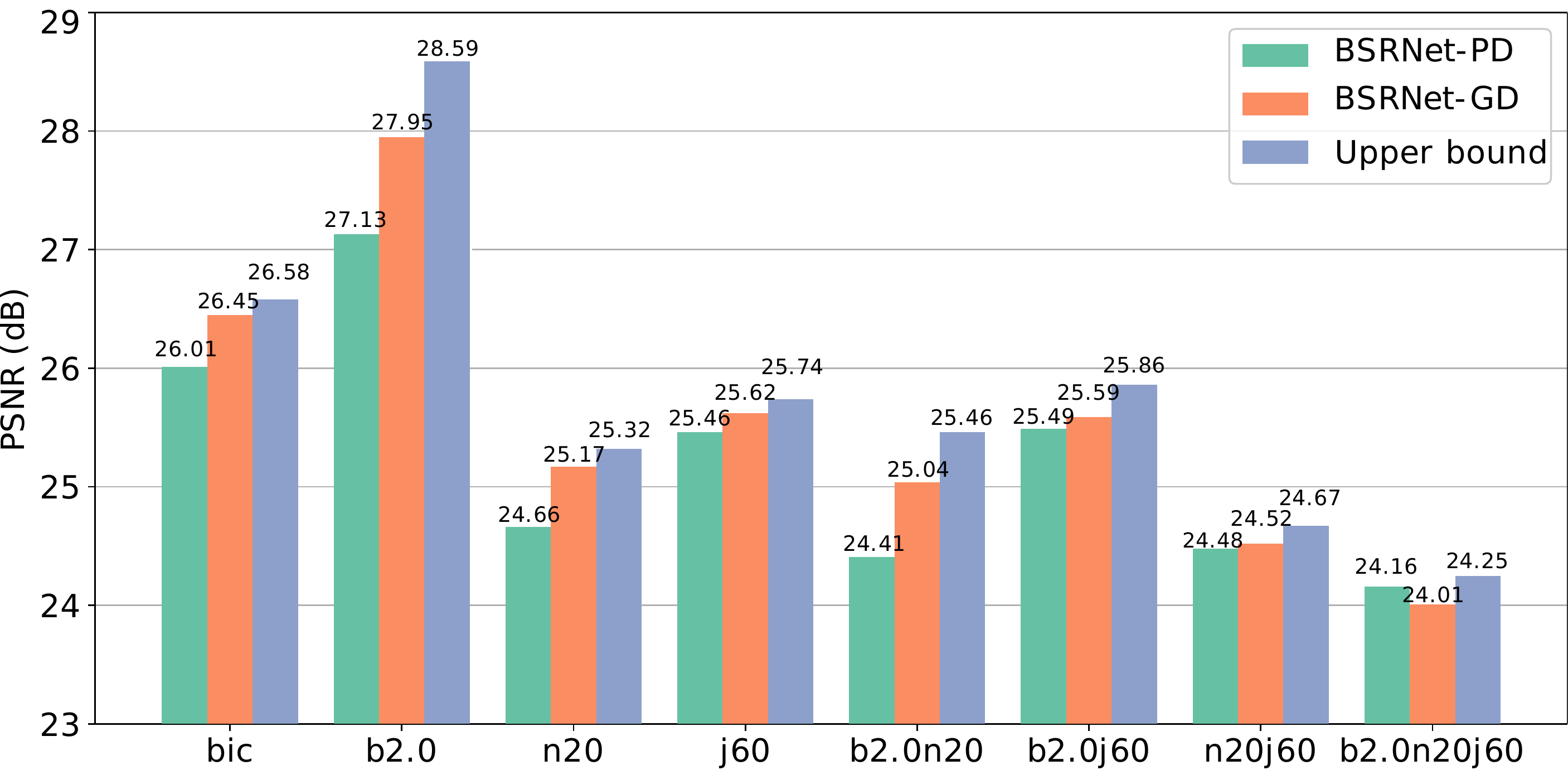}
   \caption{Comparison of PSNR (dB) of BSRNet with different degradation models.}
   \label{fig:LearningAbility}
\end{figure}
% \vspace{-0.4cm}
% The "Distance" present the distance between upper bound and BSRNet-PD/ BSRNet-GD.

\subsection{Analysis of Practical Blind SR}

\textbf{Practical Degradation Model.} We firstly train BSRNet with the PD model to get BSRNet-PD. Figure \ref{fig:LearningAbility} shows that BSRNet-PD has a significant drop on corner cases bic, blur2.0, noise20, blur2.0n20 while having a minimal drop on corner cases b2.0j60 and noise20j60.
% Specifically, the PSNR distances on validation \textit{b2.0, b2.0n20, n20} and \textit{bic} validation dataset are 1.46 dB 1.05 dB, 0.66 dB and 0.57 dB, respectively.
Interestingly, in complex case b2.0n20j60, the PSNR distance between BSRNet-PD and the upper bound is 0.09 dB, which is a tiny drop since the PD model focuses on the combination of the blur, noise, and JPEG. 
\vspace{-0.2cm}

\begin{figure}[h]
\setlength{\belowcaptionskip}{-0.4cm}

  \centering
   \includegraphics[width=1\linewidth]{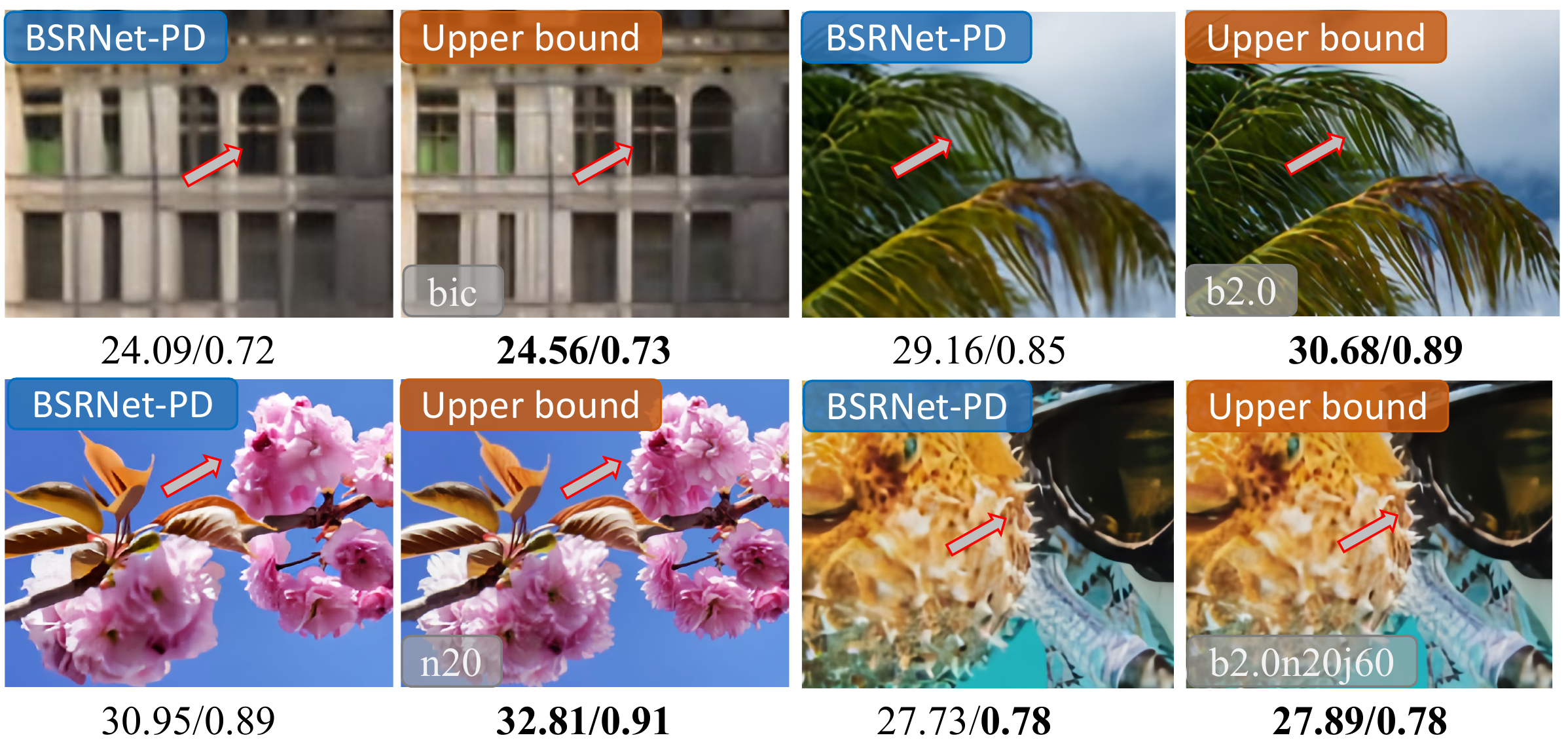}

   \caption{Visual comparisons of BSRNet-PD and the corresponding upper bound with PSNR (dB)/SSIM.}
   \label{fig: AnalysisVS1}
\end{figure}

Figure \ref{fig: AnalysisVS1} shows that BSRNet-PD fails to generate realistic textures on corner cases bic, b2.0, and n20, while the visual results on complex case b2.0n20j60 are promising compared with the upper bound. The PSNR value has a small drop of 0.16 dB, which is acceptable.

% The first comparison is in \textit{bicubic} degradation type. The structure of the window by BSRNet-PD is not clear compared with the upper bound. The similar observation is also found on the \textit{b2.0} and \textit{n20} case. The PSNR value of BSRNet-PD is obviously lower than the corresponding upper bound. Interestingly, BSRNet-PD can generate a similar realistic texture with the upper bound on b2.0n20j60. From the last comparison of Figure \ref{fig: AnalysisVS1}, the fish's texture is as clear as the texture in 
% upper bound. The PSNR distance from 27.73 of BSRNet to 27.89 of the upper bound is a small drop, which is acceptable.     
The analysis for the PD model presents three crucial points: \textbf{1)} The blind SR Network can handle the most complex case well.
\textbf{2)} The performance of the blind SR network  have a slight drop on a few corner cases, such as n20j60. \textbf{3)} The quantitative and visual results of most corner degradation types have a significant drop compared with the upper bound.    

\textbf{Gated Degradation Model.} 
To address the \textit{\textbf{issue 1}} described in Section \ref{section:inro}, we apply the proposed GD model to generate all combinations of degradation types for the BSRNet (named BSRNet-GD). Interestingly, Figure \ref{fig:LearningAbility} shows that BSRNet-GD achieves 0.82 dB and 0.63 dB improvement on the corner case b2.0 and b2.0n20, respectively. The performance of other corner cases is closer to the corresponding upper bound. The PSNR value of BSRNet-GD has a slight drop by \textbf{0.13 dB} on complex case b2.0n20j60 compared with BSRNet-PD. These results support the \textit{\textbf{solutions of issue 2} } in Section  \ref{section:inro}. A blind SR network with our proposed GD model can surprisingly achieve significant performance on all degradation cases.

\vspace{-0.15 cm}

% Although the PSNR distance of \textit{b2.0} and \textit{b2.0n20} is 0.64dB and 0.42 dB compared with upper bound,

\begin{figure}[h]
\setlength{\belowcaptionskip}{-0.25cm}
  \centering
   \includegraphics[width=1\linewidth]{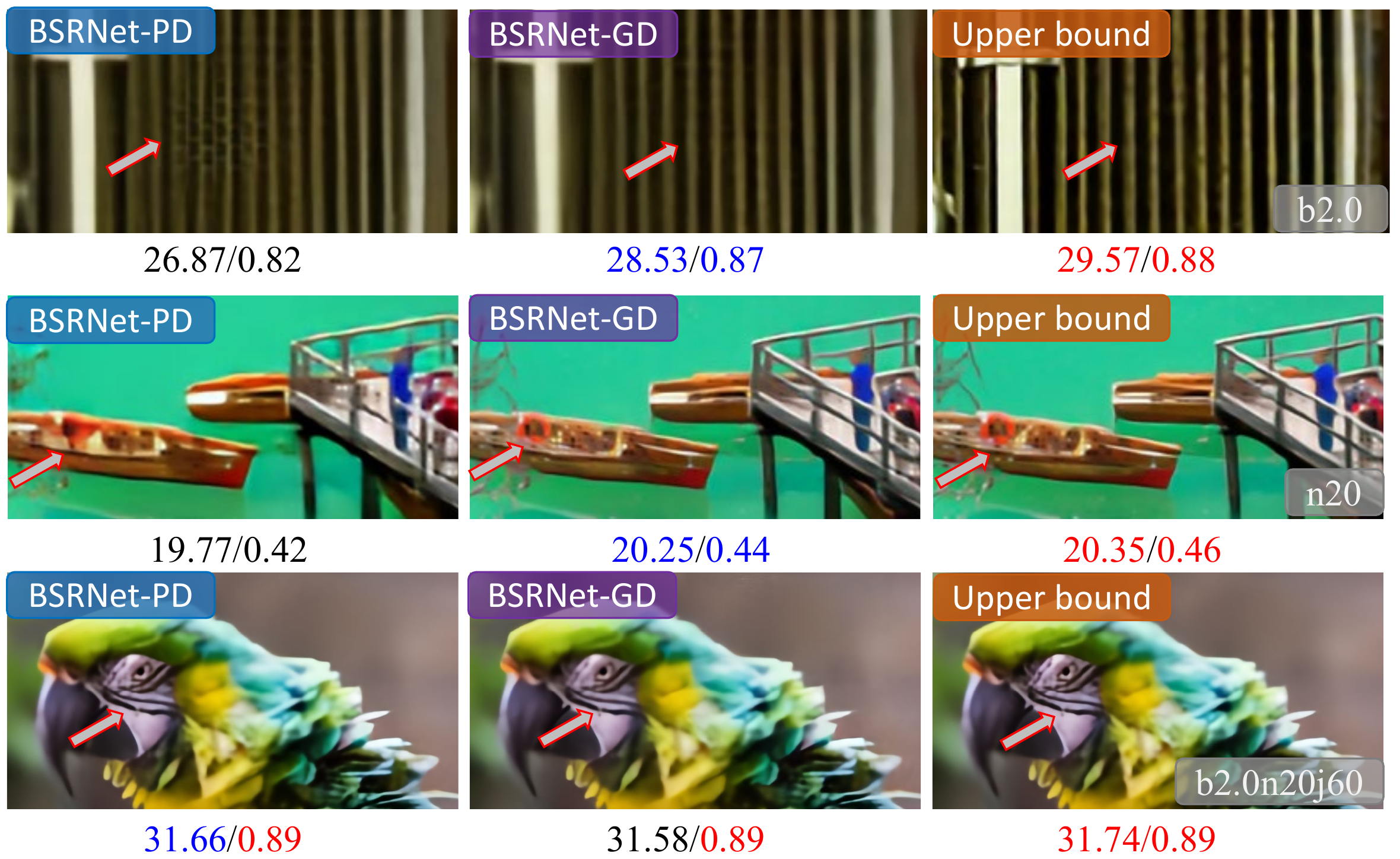}

   \caption{Visual comparisons of  BSRNet-PD, BSRNet-GD, and the corresponding upper bound with PSNR (dB)/SSIM.}
   \label{fig: AnalysisVS2}
\end{figure}

Figure \ref{fig: AnalysisVS2} shows that the BSRNet-GD can generate more realistic textures than BSRNet-PD on b2.0 and n20 degradation. The sacrifice of complex case b2.0n20j60 is completely acceptable because we can hardly tell the difference between BSRNet-PD and BSRNet-GD on visual results. Although the practical degradation model can handle some special cases, it is obvious that the practical degradation model cannot guarantee promising quantitative and qualitative results in all corner cases. These quantitative and qualitative comparisons confirm the effectiveness of upper bounds on the \textit{\textbf{issues 3}} described in Section \ref{section:inro}.    

Based on the proposed gated degradation model: \textbf{1)} A blind SR network has a tiny sacrifice in the complex case. \textbf{2)} The performance of corner cases can achieve obvious improvement compared with the PD model. \textbf{3)} A blind SR network can handle all of the degradation types with a small performance drop compared with the upper bound.

%% file: latex/tabel_analysis.tex
% \usepackage{multirow}

\begin{table}[h]
\centering
\caption{Average PSNR (dB) of BSRNet with classical degradation models in $\times$4 blind SR.}
\label{tab: classicalblindSR}
\resizebox{0.47\textwidth}{!}{
\begin{tabular}{cccccc}
\hline
\multirow{2}{*}{Method} & \multicolumn{5}{c}{Blur degradation types} \\ \cline{2-6} 
 & bic & 0.6 & 1.2 & 1.8 & 2.4 \\ \hline
BSRNet-FAIG \cite{xie2021finding}) & 26.51  & 27.25 & 28.07 & 28.42 & 28.43 \\
Upper bound & 26.75 & 27.46 & 28.43 & 28.71 & 28.74 \\ \hline
\end{tabular}}
\end{table}

%% file: latex/Experiments.tex
\section{Experiments} % p5.5-p8
\subsection{Datasets and Implementation Details}
\label{section: ExperimentSetting}
\textbf{Datasets.} Following existing blind SR methods \cite{zhang2018learning,gu2019blind,luo2020unfolding,wang2021unsupervised,wang2021real,zhang2021designing}, we use DIV2K (800 images)\cite{agustsson2017ntire} and Flickr2K (2650 images) \cite{timofte2017ntire} dataset for training. The training images are randomly cropped to 128$\times$128 patches that are blurred, noised, and compressed (JPEG). We use benchmark datasets BSD100 \cite{martin2001database} and Urban100 \cite{huang2015single} for evaluation. 

\textbf{Degradation model.} 
To ensure fair quantitative comparison, we adopt a light degradation model to generate the dataset. Following the setting of BSRGAN\cite{zhang2021designing} and RealESRGAN\cite{wang2021real}, the light degradation model includes isotropic Gaussian blur [0.1, 3.0], additive Gaussian noise [1, 30], and JPEG [40, 95]. The down-sampling adopts $\times$4 bicubic in the RealESRGAN version. For the proposed GD model, the probability of every gate is set to 0.5 to generate all degradation cases.

\textbf{Baselines.} Based on the analysis in Section \ref{sec:analysis}, the proposed GD model is applied to the representative networks as our proposed baseline network. We employ RRDBNet \cite{wang2021real,zhang2021designing} and SwinIR \cite{liang2021swinir} to get the baseline networks: CNN-based RRDBNet-GD, transformer-based SwinIR-GD, and GAN-based baseline BSRGAN-GD and SwinIRGAN-GD.   

\input{latex/table1}

\textbf{\textit{Practical8.}} In order to quantitatively  conduct evaluation, we propose \textit{Practical8} test dataset to evaluate blind SR methods.
\textit{Practical8} consists of $\{$bic, b2.0, n20, j60, b2.0n20, b2.0j60, n20j60, b2.0n20j60$\}$. The degradation types in \textit{Practical8} are based on the combinations of degradation types in the training dataset. The evaluation metric employs PSNR to compare MSE-based methods and PSNR/NIQE for GAN-based methods.

\textbf{Training.} In our experiments, the Adam\cite{kingma2014adam} optimization method with $ \beta_1=0.9 $ and $ \beta_1=0.99$ is used for training. The initial learning rate is set to $2 \times 10^{-4} $, which is reduced by a half for multi-step $ [25\times10^4, 50\times10^4, 75\times10^4, 100\times10^4]$. A total of $ 100\times10^4$ iterations are executed by PyTorch. The loss function adopts L1 loss between SR results and HR images.

\begin{figure*}[ht]

\setlength{\belowcaptionskip}{-0.38 cm}

  \centering
   \includegraphics[width=1\linewidth]{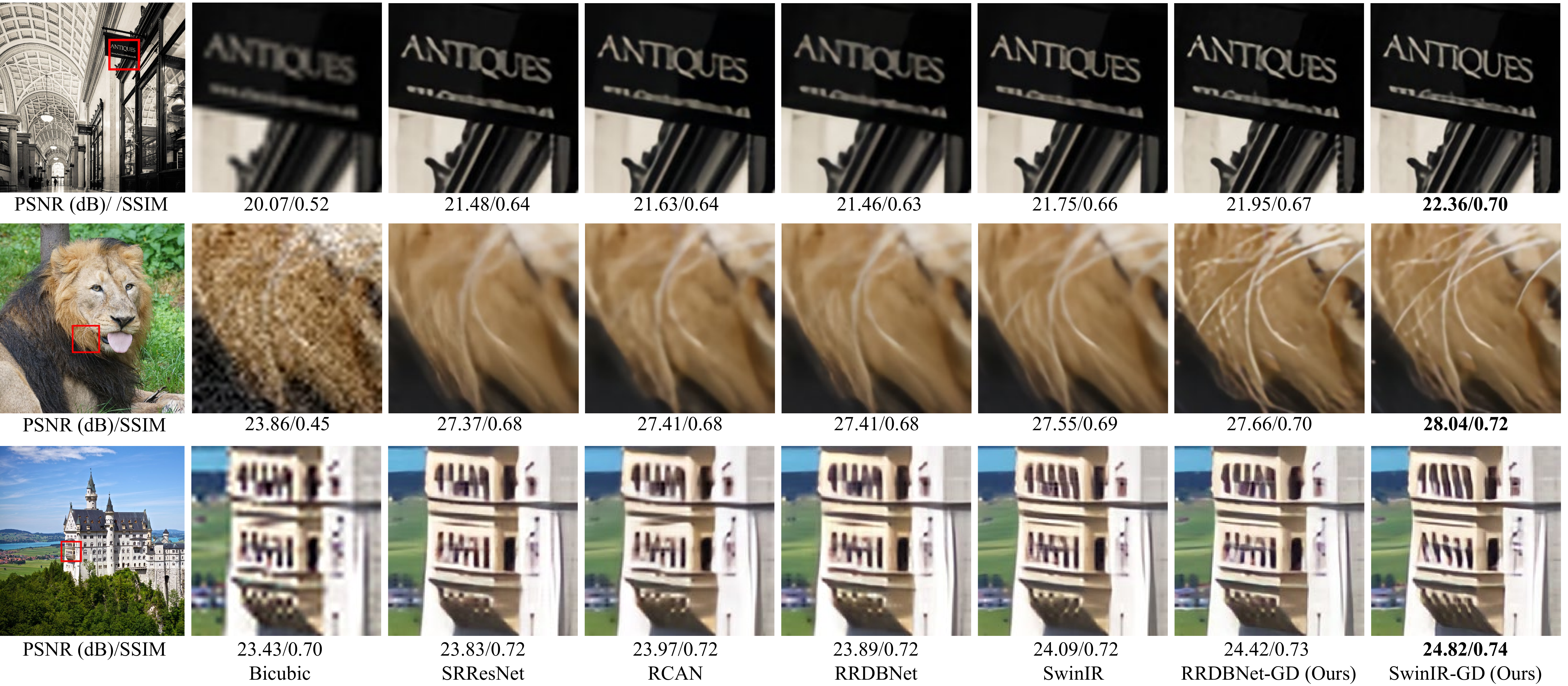}

   \caption{Visual comparisons of our methods and others in $\times$4 super-resolution. Please zoom in for a better view.}
   \label{fig: VisualResults1}
\end{figure*}

\subsection{Experiments on MSE-based blind SR}
\label{sec:experiments1}

% We analyze the performance of the proposed method with multiple SOTA methods. FAIG\cite{xie2021finding} proposed a blind MSRResNet which is comparable with the SOTA methods DAN\cite{luo2020unfolding} and DASR\cite{wang2021unsupervised}. RRDBNet is adopted by the SOTA practical blind SR methods BSRGAN\cite{zhang2021designing} and Real-ESRGAN\cite{wang2021real}. SwinIR\cite{liang2021swinir} achieve a great improvement on SR problem based on a powerful backbone Swin-transformer\cite{liu2021swin}. Therefore, the MSRResNet, RRDBNet and SwinIR are the representative models in blind SR. In addition, the representative non-blind SR method RCAN\cite{zhang2018rcan} is also be considered as a basic method.   

% Current prevalent blind SR methods include IKC\cite{gu2019blind}, DAN\cite{luo2020unfolding}, and DASR\cite{wang2021unsupervised}. 

% Recently, FAIG \cite{xie2021finding} directly employs one-branch SR network to achieve SOTA results compared with existing classical two-branch blind SR methods, such as DAN\cite{luo2020unfolding} and DASR\cite{wang2021unsupervised}. 

\textbf{Networks.} 
Here we consider a series of representative networks for quantitative comparisons, such as SRResNet which is used in prevalent kernel estimation methods, \cite{gu2019blind,luo2020unfolding,bell2019blind, Liang_2021_ICCV,liang2021flow} and RRDB network which is used in BSRGAN\cite{zhang2021designing} and RealESRGAN\cite{wang2021real} to handle practical blind SR. In addition, the representative network RCAN\cite{zhang2018rcan} and SwinIR \cite{liang2021swinir} are also employed for quantitative comparison. Notably, all networks are adjusted to the same setting and parameter level to ensure a fair comparison.     

\input{latex/table2}

\textbf{Comparison with the state-of-the-art.} Table \ref{tab: PDvsGD} shows the quantitative comparisons. Firstly, we find that RRDBNet is only about 0.03 dB higher than SRResNet-FAIG on PSNR. Interestingly, the non-blind SR method RCAN achieves better performance than SRResNet-FAIG and RRDBNet. Benefit from channel attention design, RCAN outperforms RRDBNet by about 0.1 dB on all corner degradations in \textit{Practical8}. Furthermore, SwinIR achieves the highest performance compared with other methods. 
% Indeed, RRDBNet is proposed for GAN-based SR, which maybe can generate more detailed textures by incorporating the feature of different convolution layers.
Secondly, the average performance of the proposed baseline RRDBNet-GD and SwinIR-GD achieves significant improvement (0.3-0.6 dB) on BSD100 and urban100 datasets. 
Figure \ref{fig: VisualResults1} shows that our method could generate visually pleasing results than other works.

% Figure11  shows  some  visual  examples,  where  we  observe  that our method could generate more realistic textures withoutintroducing additional artifacts (please see the windows inImg 233 and feathers in Img 242).In  addition,  we  use  DIV2K  and  Flickr2K  validation  set(as  illustrated  in  Section  5.1)  to  evaluate  the  differencebetween  GT  images  and  GAN  outputs.  The  NIQE  scoredistribution  is  shown  in  Figure  9.  We  observe  that  mostof  GAN-based  methods  can  achieve  lower  NIQE  scoresthan  GT  images.  Specifically,  for  SRGAN,  ESRGAN  andRankSRGAN,  the  proportion  of  their  NIEQ  scores  lowerthan  GT  images  is76.2%,86.6%and90.4%,  respectively.The  average  NIQE  score  of  GT,  SRGAN,  ESRGAN  andRankSRGAN is 4.90, 4.12, 3.81 and 3.74, respectively. (LowerNIQE indicates better.) RankSRGAN achieves the best aver-age NIQE score than the GT images.As  the  results  may  vary  across  different  iterations,  wefurther  show  the  convergence  curves  of  RankSRGAN  inFigure  12.  Their  performance  on  NIQE  and  PSNR  are  rel-atively  stable  during  the  training  process.  For  PSNR,  theyobtain  comparable  results.  But  for  NIQE,  RankSRGAN  isconsistently better than SRGAN by a large margin

\textbf{Upper bound.}
 To further evaluate the performance of the blind SR networks, we train 8 SR models with the specific degradation types in \textit{\textbf{Practical8}}. Since RRDBNet is adopted in SOTA practical blind SR methods BSRGAN\cite{zhang2021designing} and RealESRGAN\cite{wang2021real}, RRDBNet is selected as the basic network to obtain the upper bound. 
%  Specifically, the upper bound of ``bic'' degradation represents a special SR network on the ``bicubic'' training dataset. Another upper bound is set to the same setting.
 Table \ref{tab: PDvsGD} shows that blind SR networks with a practical degradation model have a significant drop compared with the upper bound on some corner degradation cases. The most interesting aspect is that the quantitative difference between a specific case and the upper bound is very large. Take RRDBNet as an example, it is apparent that the bic, b2.0, and b2.0n20 cases have a large performance drop compared with other cases.  
 
Based on the proposed GD model, there is a significant improvement in all corner cases, such as bicubic (non-blind SR), b2.0 (classical blind SR), and complex case b2.0n20j60 (practical blind SR). Notably, there are also great differences in the improvement of different cases. For example, the b2.0j60 case has the smallest improvement, and bic case has a great improvement compared with the upper bounds.

\textbf{Network capacity.}
Table \ref{tab: parameter} shows the comparisons of blind SR networks with different network parameters and structures on the GD model. The SRResNet-16 with 16 residual blocks has 1.52M parameters, but it just has a 0.39 dB drop compared with the upper bound on average PSNR. Furthermore, SRResNet-46 and RRDBNet-5 get about 0.1 dB improvement compared with SRResNet-16. Benefitting from the attention mechanism, RCAN and SwinIR-v1 (version1) achieve better performance with similar parameters. Finally, SwinIR-v2 (version2) with 11.9 M can further improve the SR results. Interestingly, the degradation b2.0n20j60 only has a slight improvement (0.07 dB), while the easy corner degradations have a significant improvement (e.g., 0.27 dB in bicubic).    

  \input{latex/table3}

\textbf{Light vs. hard degradation models.}
We further apply the proposed GD model with a hard scenario, which includes various blur types (isotropic, anisotropic, generalized isotropic/anisotropic, and plateau isotropic/an-isotropic Gaussian blur), noises (additive grey/color Gaussian noise and Poisson grey/color noise) and JPEG compression. Table \ref{tab: LightvsHard} shows that the performance of RRDBNet-GD-hard has a slight drop on the light cases, while it has a more significant improvement on the new cases.

\input{latex/table4}

\begin{figure*}[ht]
\setlength{\abovecaptionskip}{-0.01 cm}

\setlength{\belowcaptionskip}{-0.2 cm}

  \centering
   \includegraphics[width=0.95\linewidth]{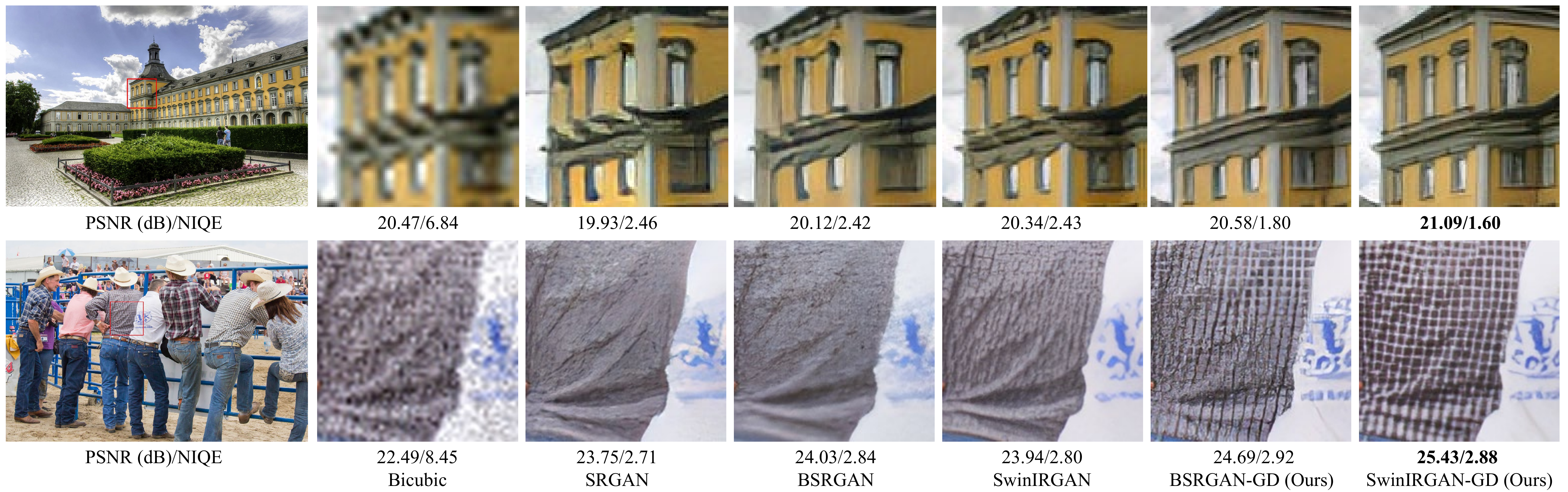}

   \caption{Visual comparisons of our methods and others in $\times$4 Blind SR. Lower NIQE score indicates better perceptual quality, and higher PSNR indicates less distortion. Please zoom in for a better view.}
   \label{fig: VisualResults2}
\end{figure*}

\subsection{Experiments on GAN-based Blind SR}
\textbf{Experimental setup.}
Similar to Section \ref{section: ExperimentSetting}, we adopt the same settings to train the GAN-based networks. We train three representative models, SRGAN \cite{ledig2017photo}, BSRGAN\cite{zhang2021designing} (RealESRGAN\cite{wang2021real}), and SwinIRGAN \cite{liang2021swinir} by the same light degradation model in Section \ref{section: ExperimentSetting}. The loss function combines L1 loss, perceptual loss, and GAN loss, with weights [1, 1, 0.1], respectively. The baseline BSRGAN-GD and SwinIRGAN-GD are trained on the proposed GD model. The discriminator adopts a U-Net structure in RealESRGAN \cite{wang2021real}. The upper bound of \textit{Practical8} is also provided to evaluate the performance of different GAN models quantitatively.   

\textbf{Comparison with the stat-of-the-arts.} Table \ref{tab: gan} shows SRGAN tends to sacrifice PSNR performance to generate perceptual textures while BSRGAN and SwinIRGAN can achieve higher reconstructive performance when generating texture details. Based on our proposed GD model, the reconstructive performance achieves further improvement compared with the practical degradation model. Interestingly, SwinIRGAN pays more attention to reconstruction performance PSNR while the perceptual metric NIQE value is higher than BSRGAN-GD. Figure \ref{fig: VisualResults2} shows that our methods can generate realistic visual results compared with existing methods. 
We validate our method on a real-world dataset RealSRSet used in BSRGAN \cite{zhang2021designing}, which consists of real images downloaded from the Internet. BSRGAN-GD achieves 5.11 in NIQE, much better than BSRGAN with a light practical degradation model, which is 6.06.

\subsection{Discussion}
To further adapt the proposed GD model for the real-world scenario, an intuitive way is to enlarge the degradation space. We apply this simple scheme in Section \ref{sec:experiments1}, Table \ref{tab: LightvsHard} shows the potential ability to handle complex cases in real-world scenarios. Different from BSRGAN\cite{zhang2021designing} and RealESRGAN\cite{wang2021real} that tends to provide a powerful degradation model, our work focuses on how to fairly and quantitatively evaluate blind SR networks. In addition, our proposed degradation model is complement to that of BSRGAN and addresses the important corner cases that were not considered in BSRGAN. We can easily apply the proposed GD strategy to a complex degradation model, such as the degradation model in BSRGAN and RealESRGAN. In summary, as the degradations are extremely complex in real-world applications, the degradation model, baseline, and upper bound would be an important topic for future blind SR research.

% We can observe how close is a blind SR method to the upper bound in a specific degradation type.

% In addition, we can easily apply the proposed GD strategy to any degradation model, such as degradation model with shuffle in BSRGAN and high order in RealESRGAN. 
% 
% Our paper focuses on how to fairly evaluate blind SR
% networks. To this end, we investigate importance issues
% including degradation models, baselines, and performance
% upper bounds. Our proposed degradation model is complement
% to that of BSRGAN and addresses the important
% corner cases that were not considered in BSRGAN. Based
% on the proposed degradation model, we conduct a fair and
% quantitative evaluation of popular blind SR networks under
% the same setting, which demonstrates the high effectiveness
% of our degradation model and provides strong baselines and
% performance upper bounds as useful references for future
% research.

% \vspace{0.3 cm}

%% file: latex/table1.tex
% \usepackage{multirow}
% \usepackage{graphicx}

\begin{table*}[ht]
\centering
\caption{Average PSNR (dB) of different methods in $\times$4 blind SR on BSD100 \cite{martin2001database} and Urban100 \cite{huang2015single}. The top two results are highlighted in red and blue, respectively. Note that we ensure all methods have similar model size for a fair comparison.
%we ethe parameters of all networks are consistent, thus the comparison is fair.
}
\label{tab: PDvsGD}
\resizebox{\textwidth}{!}{%
\begin{tabular}{ccccccccccc}
\hline
\multirow{2}{*}{Dataset} & \multirow{2}{*}{Method} & \multicolumn{9}{c}{Degradation Types} \\ \cline{3-11} 
 &  & bic & b2.0 & n20 & j60 & b2.0n20 & b2.0j60 & n20j60 & \multicolumn{1}{c|}{b2.0n20j60} & Average \\ \hline \hline
% \multirow{7}{*}{Set14} & Bicubic & 25.00 & 25.34 & 21.77 & 24.29 & 21.91 & 24.51 & 21.46 & \multicolumn{1}{c|}{21.73} & 23.25 \\
%  & RCAN \cite{zhang2018rcan} & 26.11 & 27.44 & 24.79 & 25.61 & 24.58 & 25.75 & 24.62 & \multicolumn{1}{c|}{24.30} & 25.40 \\

%  & SRResNet \cite{xie2021finding} & 26.12 & 27.39 & 24.71 & 25.56 & 24.41 & 25.61 & 24.55 & \multicolumn{1}{c|}{24.16} & 25.31 \\
%  & RRDBNet \cite{wang2021real,zhang2021designing} & 26.06 & 27.48 & 24.70 & 25.56 & 24.46 & 25.66 & 24.56 & \multicolumn{1}{c|}{24.21} & 25.34 \\
%  & SwinIR\cite{liang2021swinir} & 26.44 & 27.78 & 25.04 & 25.78 & 24.73 & 25.89 & 24.75 & \multicolumn{1}{c|}{24.37} & 25.60 \\ \cdashline{2-11} 
%  & RRDBNet+GD (ours) & 26.53 & 28.25 & 25.28 & 25.68 & 25.22 & 25.62 & 24.59 & \multicolumn{1}{c|}{24.15} & 25.67 \\
%  & SwinIR+GD (ours) & 26.91 & 28.61 & 25.66 & 25.81 & 25.73 & 25.77 & 24.76 & \multicolumn{1}{c|}{24.30} & 25.94 \\ \cdashline{2-11} 
%  & Upper-bound & 26.75 & 28.74 & 25.48 & 25.81 & 25.63 & 25.96 & 24.74 & \multicolumn{1}{c|}{24.32} & 25.93 \\ \hline \hline
\multirow{7}{*}{BSD100} & Bicubic & 24.63 & 25.40 & 21.56 & 24.06 & 21.90 & 24.65 & 21.22 & \multicolumn{1}{c|}{21.72} & 23.14 \\
 & RCAN\cite{zhang2018rcan} & 25.65 & 26.77 & 24.63 & 25.16 & 24.39 & 25.36 & 24.36 & \multicolumn{1}{c|}{24.15} & 25.06 \\

 & SRResNet-FAIG \cite{xie2021finding} & 25.58 & 26.72 & 24.53 & 25.11 & 24.26 & 25.29 & 24.32 & \multicolumn{1}{c|}{24.07} & 24.99 \\
 & RRDBNet\cite{wang2021real,zhang2021designing} & 25.62 & 26.76 & 24.58 & 25.13 & 24.33 & 25.32 & 24.34 & \multicolumn{1}{c|}{24.11} & 25.02 \\
 & SwinIR \cite{liang2021swinir} & 25.84 & 27.05 & 24.77 & \textcolor{blue}{25.27} & 24.48 & \textcolor{red}{25.44} & \textcolor{red}{24.44} & \multicolumn{1}{c|}{\textcolor{red}{24.18}} & 25.18 \\ \cdashline{2-11} 
 & RRDBNet-GD (ours) & \textcolor{blue}{26.25} & \textcolor{blue}{27.31} & \textcolor{blue}{25.31} & 25.23 & \textcolor{blue}{24.95} & 25.32 & \textcolor{blue}{24.38} & \multicolumn{1}{c|}{24.07} & \textcolor{blue}{25.35} \\
 & SwinIR-GD (ours) & \textcolor{red}{26.61} & \textcolor{red}{27.58} & \textcolor{red}{25.64} & \textcolor{red}{25.30} & \textcolor{red}{25.30} & \textcolor{blue}{25.39} & \textcolor{red}{24.44} & \multicolumn{1}{c|}{\textcolor{blue}{24.14}} & \textcolor{red}{25.55} \\ \cdashline{2-11}

 & Upper bound (RRDBNet) & 26.36 & 27.68 & 25.46 & 25.30 & 25.34 & 25.49 & 24.45 & \multicolumn{1}{c|}{24.15} & 25.53 \\ \hline \hline

\multirow{7}{*}{Urban100} & Bicubic & 21.89 & 22.54 & 20.00 & 21.50 & 20.36 & 22.02 & 19.74 & \multicolumn{1}{c|}{20.20} & 21.03 \\
 & RCAN \cite{zhang2018rcan} & 23.65 & 24.67 & 22.93 & 23.35 & 22.59 & 23.36 & 22.77 & \multicolumn{1}{c|}{22.35} & 23.21 \\

 & SRResNet-FAIG \cite{xie2021finding} & 23.54 & 24.42 & 22.88 & 23.26 & 22.42 & 23.16 & 22.73 & \multicolumn{1}{c|}{22.19} & 23.08 \\
 & RRDBNet\cite{wang2021real,zhang2021designing} & 23.53 & 24.46 & 22.89 & 23.28 & 22.48 & 23.17 & 22.75 & \multicolumn{1}{c|}{22.24} & 23.10 \\
 & SwinIR \cite{liang2021swinir} & 24.16 & 25.10 & 23.34 & \textcolor{blue}{23.73} & 22.86 & \textcolor{red}{23.62} & \textcolor{blue}{23.09} & \multicolumn{1}{c|}{\textcolor{red}{22.53}} & \textcolor{blue}{23.55} \\ \cdashline{2-11} 
 & RRDBNet-GD (ours) & \textcolor{blue}{24.51} & \textcolor{blue}{25.39} & \textcolor{blue}{23.57} & 23.67 & \textcolor{blue}{23.05} & 23.18 & 22.92 & \multicolumn{1}{c|}{22.13} & \textcolor{blue}{23.55} \\
 & SwinIR-GD (ours) & \textcolor{red}{25.55} & \textcolor{red}{26.12} & \textcolor{red}{24.40} & \textcolor{red}{24.11} & \textcolor{red}{23.83} & \textcolor{blue}{23.56} & \textcolor{red}{23.26} & \multicolumn{1}{c|}{\textcolor{blue}{22.42}} & \textcolor{red}{24.16} \\ \cdashline{2-11} 
 & Upper bound (RRDBNet) & 25.13 & 26.38 & 23.91 & 23.97 & 23.56 & 23.62 & 23.18 & \multicolumn{1}{c|}{22.44} & 24.02 \\ \hline
\end{tabular}%
}
\end{table*}

%% file: latex/table2.tex
% \usepackage{multirow}
% \usepackage{graphicx}
\begin{table*}[ht]
\centering
\caption{Average PSNR (dB) of networks of different capacity in $\times$4 blind SR with the proposed gated degradation model on Set14 \cite{zeyde2010single}. }
\label{tab: parameter}
\resizebox{\textwidth}{!}{%
\begin{tabular}{ccccccccccc}
\hline
\multirow{2}{*}{Method} & \multirow{2}{*}{\begin{tabular}[c]{@{}c@{}}\#Para.\\ (M)\end{tabular}} & \multicolumn{9}{c}{Degradation Types} \\ \cline{3-11} 
 &  & bic & b2.0 & n20 & j60 & b2.0n20 & b2.0j60 & n20j60 & \multicolumn{1}{c|}{b2.0n20j60} & Average \\ \hline \hline
Bicubic & - & 25.00 & 25.34 & 21.77 & 24.29 & 21.91 & 24.51 & 21.46 & \multicolumn{1}{c|}{21.73} & 23.25 \\
SRResNet-16 & 1.52 & 26.45 & 27.94 & 25.17 & 25.59 & 25.04 & 25.56 & 24.53 & \multicolumn{1}{c|}{24.04} & 25.54 \\
SRResNet-46 & 3.73 & 26.49 & 28.16 & 25.23 & 25.67 & 25.12 & 25.57 & 24.58 & \multicolumn{1}{c|}{24.09} & 25.61 \\
RCAN & 3.87 & 26.62 & 28.31 & 25.36  & 25.75  & 25.33  & 25.66  & 24.68 & \multicolumn{1}{c|}{24.19} & 25.74 \\
RRDBNet-5 & 3.75 & 26.53 & 28.25 & 25.28 & 25.68 & 25.22 & 25.62 & 24.59 & \multicolumn{1}{c|}{24.15} & 25.67 \\
SwinIR-v1 & 3.85 & \textcolor{blue}{26.94} & \textcolor{blue}{28.59} & \textcolor{blue}{25.67} & \textcolor{blue}{25.83} & \textcolor{blue}{25.73} & \textcolor{blue}{25.77} & \textcolor{blue}{24.77} & \multicolumn{1}{c|}{\textcolor{blue}{24.30}} & \textcolor{blue}{25.95} \\
SwinIR-v2 & 11.90 & \textcolor{red}
{27.21} & \textcolor{red}
{28.84} & \textcolor{red}
{25.92} & \textcolor{red}
{26.07} & \textcolor{red}
{25.87} & \textcolor{red}
{25.87} & \textcolor{red}
{24.91} & \multicolumn{1}{c|}{\textcolor{red}
{24.37}} & \textcolor{red}
{26.13} \\ \hdashline

\begin{tabular}[c]{@{}c@{}}Upper bound \\ (RRDBNet-5)\end{tabular} & 3.75 & 26.75 & 28.74 & 25.48 & 25.81 & 25.63 & 25.96 & 24.74 & \multicolumn{1}{c|}{24.32} & 25.93 \\ \hline
\end{tabular}%
}
\end{table*}

%% file: latex/table3.tex
% \usepackage{multirow}

\begin{table}[]
\centering
\caption{Average PSNR (dB) of RRDBNet in $\times$4 blind SR with light and hard degradation models on Set14 \cite{zeyde2010single}.}
\label{tab: LightvsHard}
\resizebox{0.47\textwidth}{!}{
\begin{tabular}{ccccc}
\hline
\multirow{2}{*}{Method} & \multicolumn{4}{c}{Degradation Types} \\ \cline{2-5} 
 & bic & b2.0 & n20 & j60 \\ \hline
RRDBNet-GD-light & \textbf{26.53} & \textbf{28.25} & \textbf{25.28} & \textbf{25.68} \\
RRDBNet-GD-hard & \textbf{26.53} & 28.11 & 25.22 & \textbf{25.68} \\ \hline

\multicolumn{1}{l}{} & b2.0n20 & b2.0j60 & n20j60 & b2.0n20j60 \\ \hline

RRDBNet-GD-light & \textbf{25.22} & 25.62 & \textbf{24.59} & \textbf{24.15} \\
RRDBNet-GD-hard & 25.12 &\textbf{ 26.65} & 24.56 & 24.08 \\ \hline

\multicolumn{1}{l}{} & color-n20 & Poisson-n20 
& 
\begin{minipage}[b]{0.1\columnwidth}
		\centering
		\raisebox{-.5\height}{\includegraphics[width=\linewidth]{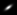}}
	\end{minipage}
& \begin{minipage}[b]{0.1\columnwidth}
		\centering
		\raisebox{-.5\height}{\includegraphics[width=\linewidth]{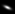}}
	\end{minipage}\\ \hline

RRDBNet-GD-light & 24.93 & 24.66 & 26.71 & 26.56 \\
RRDBNet-GD-hard & \textbf{25.52} & \textbf{25.43} & \textbf{26.90} & \textbf{26.83} \\ \hline

\end{tabular}}
\end{table}
% \vspace{-0.8 cm}

%% file: latex/table4.tex
% \usepackage{multirow}
\begin{table*}[ht]
\centering
\caption{Average NIQE/ PSNR (dB) of different GAN-based methods in $\times$4 blind SR on Urban100 \cite{huang2015single}. The top two results are highlighted in red and blue, respectively. Note that we ensure all methods have similar model size for a fair comparison.}

\label{tab: gan}
\resizebox{\textwidth}{!}{%
\begin{tabular}{ccccccccccc}
\hline
\multirow{2}{*}{Method} & \multirow{2}{*}{Metric} & \multicolumn{9}{c}{Degradation Types} \\ \cline{3-11} 
 &  & bic & b2.0 & n20 & j60 & b2.0n20 & b2.0j60 & n20j60 & \multicolumn{1}{c|}{b2.0n20j60} & Average \\ \hline \hline
\multirow{2}{*}{Bicubic} & NIQE & 7.08 & 7.89 & 8.97 & 7.35 & 8.42 & 7.93 & 8.99 & \multicolumn{1}{c|}{8.37} & 8.13 \\
 & PSNR & 21.89 & 22.54 & 20.00 & 21.50 & 20.36 & 22.02 & 19.74 & \multicolumn{1}{c|}{20.20} & 21.03 \\
\multirow{2}{*}{SRGAN\cite{ledig2017photo}} & NIQE & 4.25 & 5.00 & \textcolor{red}{3.49} & \textcolor{red}{3.88} & \textcolor{red}{3.69} & \textcolor{blue}{4.59} & \textcolor{red}{3.46} & \multicolumn{1}{c|}{\textcolor{red}{3.65}} & \textcolor{red}{4.00} \\
 & PSNR & 21.75 & 23.16 & 21.08 & 21.55 & 21.68 & 22.42 & 20.95 & \multicolumn{1}{c|}{21.45} & 21.76 \\
 
\multirow{2}{*}{BSRGAN\cite{wang2021real,zhang2021designing}} & NIQE & 4.51 & 5.77 & 4.02 & 4.25 & 4.24 & 5.26 & 3.97 & \multicolumn{1}{c|}{\textcolor{blue}{4.36}} & 4.55 \\
 & PSNR & 22.18 & 23.39 & 21.58 & 21.96 & 21.81 & 22.51 & 21.38 & \multicolumn{1}{c|}{21.51} & 22.04 \\
 
\multirow{2}{*}{SwinIRGAN\cite{liang2021swinir}} 
 & NIQE & 4.39 & 5.01 & 4.29 & 4.40 & 4.46 & 4.91 & 4.08 & \multicolumn{1}{c|}{\textcolor{blue}{4.36}} & 4.49 \\
 & PSNR & 22.92 & 24.10 & 22.10 & \textcolor{blue}{22.48} & 22.18 & \textcolor{blue}{22.84} & \textcolor{blue}{21.82} & \multicolumn{1}{c|}{\textcolor{blue}{21.83}} & 22.53  \\ \hdashline

\multirow{2}{*}{BSRGAN-GD (ours)} & NIQE & \textcolor{blue}{4.04} & \textcolor{red}{4.27} & \textcolor{blue}{3.91} & \textcolor{blue}{3.95} & \textcolor{blue}{4.18} & 4.91 & \textcolor{blue}{3.63} & \multicolumn{1}{c|}{4.57} & \textcolor{blue}{4.18} \\
 
 & PSNR & \textcolor{blue}{23.31} & \textcolor{blue}{24.43} & \textcolor{blue}{22.51} & 22.45 & \textcolor{blue}{22.40} & 22.69 & 21.62 & \multicolumn{1}{c|}{21.62} & \textcolor{blue}{22.63} \\
 
\multirow{2}{*}{SwinIRGAN-GD (ours)} & NIQE &\textcolor{red}{ 4.01} & \textcolor{blue}{4.38} & 4.11 & 4.16 & 4.29 & \textcolor{red}{4.55} & 4.09 & \multicolumn{1}{c|}{4.72} & 4.29 \\

 & PSNR & \textcolor{red}{24.24} & \textcolor{red}{25.20} & \textcolor{red}{23.28} & \textcolor{red}{22.98} & \textcolor{red}{23.13} & \textcolor{red}{22.94} & \textcolor{red}{22.17} & \multicolumn{1}{c|}{\textcolor{red}{21.86}} & \textcolor{red}{23.23} \\ \hdashline

\multirow{2}{*}{Upper bound (BSRGAN)} & NIQE & 3.79 & 4.10 & 3.88 & 3.92 & 3.86 & 4.00 & 3.73 & \multicolumn{1}{c|}{3.87} & 3.89 \\
 & PSNR & 23.66 & 25.17 & 22.58 & 22.58 & 22.41 & 22.51 & 21.77 & \multicolumn{1}{c|}{21.52} & 22.78 \\ \hline
\end{tabular}}
\end{table*}

%% file: latex/Conclusion.tex
\section{Conclusion}
In this paper, we have proposed a gated degradation model to unify the non-blind SR, classical SR, and practical SR. Based on the proposed degradation model, we provide a detailed, quantitative, and comprehensive analysis to evaluate the learning ability of a blind SR network with classical and practical degradation models. Further, we introduce a \textit{practical8} validation dataset to evaluate the blind SR network with the performance upper bounds quantitatively. Moreover, we establish a series of strong baselines, including CNN-based, transformer-based and GAN-based. Experimental results show that our proposed baselines achieve state-of-the-art performance for practical blind SR on most degradation cases, which can facilitate future research.

\textbf{Acknowledgements}
This work was supported by the Grant of DaSAIL Project P0030935 funded by PolyU/UGC and by the National Natural Science Foundation of China (61906184). 
% \vspace{-0.5 cm}

% \textbf{Limitations.} Although a blind SR network with the proposed GD model can generate visually pleasing results, it still has limitations. 1) The performance drop on hard degradation cases (e.g., blur types) is higher than on noise and JPEG types. 2) With more degradation types, the performance of the blind SR network may gradually decline. Hence, a large degradation model may need a large network. 3) The proposed baseline SwinIR-GD can achieve promising performance, but the training cost of a transformer-based network is still high. These drawbacks have a significant impact on the practical application of blind SR networks and need to be addressed in future works.
% , parameter comparisons, upper bound comparisons and ligh/hard degradation models

%% file: PaperForReview.bbl
\begin{thebibliography}{10}\itemsep=-1pt

\bibitem{agustsson2017ntire}
Eirikur Agustsson and Radu Timofte.
\newblock Ntire 2017 challenge on single image super-resolution: Dataset and
  study.
\newblock In {\em The IEEE Conference on Computer Vision and Pattern
  Recognition (CVPR) Workshops}, volume~3, page~2, 2017.

\bibitem{bell2019blind}
Sefi Bell-Kligler, Assaf Shocher, and Michal Irani.
\newblock Blind super-resolution kernel estimation using an internal-gan.
\newblock {\em Advances in Neural Information Processing Systems}, 32, 2019.

\bibitem{cai2019toward}
Jianrui Cai, Hui Zeng, Hongwei Yong, Zisheng Cao, and Lei Zhang.
\newblock Toward real-world single image super-resolution: A new benchmark and
  a new model.
\newblock In {\em Proceedings of the IEEE/CVF International Conference on
  Computer Vision}, pages 3086--3095, 2019.

\bibitem{cornillere2019blind}
Victor Cornillere, Abdelaziz Djelouah, Wang Yifan, Olga Sorkine-Hornung, and
  Christopher Schroers.
\newblock Blind image super-resolution with spatially variant degradations.
\newblock {\em ACM Transactions on Graphics (TOG)}, 38(6):1--13, 2019.

\bibitem{dong2014learning}
Chao Dong, Chen~Change Loy, Kaiming He, and Xiaoou Tang.
\newblock Learning a deep convolutional network for image super-resolution.
\newblock In {\em European conference on computer vision}, pages 184--199.
  Springer, 2014.

\bibitem{gu2019blind}
Jinjin Gu, Hannan Lu, Wangmeng Zuo, and Chao Dong.
\newblock Blind super-resolution with iterative kernel correction.
\newblock In {\em Proceedings of the IEEE/CVF Conference on Computer Vision and
  Pattern Recognition}, pages 1604--1613, 2019.

\bibitem{haris2018deep}
Muhammad Haris, Greg Shakhnarovich, and Norimichi Ukita.
\newblock Deep backprojection networks for super-resolution.
\newblock In {\em Conference on Computer Vision and Pattern Recognition}, 2018.

\bibitem{He_2019_CVPR}
Jingwen He, Chao Dong, and Yu Qiao.
\newblock Modulating image restoration with continual levels via adaptive
  feature modification layers.
\newblock In {\em The IEEE Conference on Computer Vision and Pattern
  Recognition (CVPR)}, June 2019.

\bibitem{huang2015single}
Jia-Bin Huang, Abhishek Singh, and Narendra Ahuja.
\newblock Single image super-resolution from transformed self-exemplars.
\newblock In {\em Proceedings of the IEEE conference on computer vision and
  pattern recognition}, pages 5197--5206, 2015.

\bibitem{kim2016accurate}
Jiwon Kim, Jung Kwon~Lee, and Kyoung Mu~Lee.
\newblock Accurate image super-resolution using very deep convolutional
  networks.
\newblock In {\em Proceedings of the IEEE conference on computer vision and
  pattern recognition}, pages 1646--1654, 2016.

\bibitem{kingma2014adam}
Diederik~P Kingma and Jimmy Ba.
\newblock Adam: A method for stochastic optimization.
\newblock {\em arXiv preprint arXiv:1412.6980}, 2014.

\bibitem{ledig2017photo}
Christian Ledig, Lucas Theis, Ferenc Husz{\'a}r, Jose Caballero, Andrew
  Cunningham, Alejandro Acosta, Andrew~P Aitken, Alykhan Tejani, Johannes Totz,
  Zehan Wang, et~al.
\newblock Photo-realistic single image super-resolution using a generative
  adversarial network.
\newblock In {\em CVPR}, volume~2, page~4, 2017.

\bibitem{liang2021swinir}
Jingyun Liang, Jiezhang Cao, Guolei Sun, Kai Zhang, Luc Van~Gool, and Radu
  Timofte.
\newblock Swinir: Image restoration using swin transformer.
\newblock In {\em IEEE International Conference on Computer Vision Workshops},
  2021.

\bibitem{Liang_2021_ICCV}
Jingyun Liang, Guolei Sun, Kai Zhang, Luc Van~Gool, and Radu Timofte.
\newblock Mutual affine network for spatially variant kernel estimation in
  blind image super-resolution.
\newblock In {\em Proceedings of the IEEE/CVF International Conference on
  Computer Vision (ICCV)}, pages 4096--4105, October 2021.

\bibitem{liang2021flow}
Jingyun Liang, Kai Zhang, Shuhang Gu, Luc Van~Gool, and Radu Timofte.
\newblock Flow-based kernel prior with application to blind super-resolution.
\newblock In {\em Proceedings of the IEEE/CVF Conference on Computer Vision and
  Pattern Recognition}, pages 10601--10610, 2021.

\bibitem{lim2017enhanced}
Bee Lim, Sanghyun Son, Heewon Kim, Seungjun Nah, and Kyoung~Mu Lee.
\newblock Enhanced deep residual networks for single image super-resolution.
\newblock In {\em The IEEE conference on computer vision and pattern
  recognition (CVPR) workshops}, volume~1, page~4, 2017.

\bibitem{lugmayr2019unsupervised}
Andreas Lugmayr, Martin Danelljan, and Radu Timofte.
\newblock Unsupervised learning for real-world super-resolution.
\newblock In {\em 2019 IEEE/CVF International Conference on Computer Vision
  Workshop (ICCVW)}, pages 3408--3416. IEEE, 2019.

\bibitem{luo2020unfolding}
Zhengxiong Luo, Yan Huang, Shang Li, Liang Wang, and Tieniu Tan.
\newblock Unfolding the alternating optimization for blind super resolution.
\newblock {\em arXiv preprint arXiv:2010.02631}, 2020.

\bibitem{luo2021end}
Zhengxiong Luo, Yan Huang, Shang Li, Liang Wang, and Tieniu Tan.
\newblock End-to-end alternating optimization for blind super resolution.
\newblock {\em arXiv preprint arXiv:2105.06878}, 2021.

\bibitem{martin2001database}
David Martin, Charless Fowlkes, Doron Tal, and Jitendra Malik.
\newblock A database of human segmented natural images and its application to
  evaluating segmentation algorithms and measuring ecological statistics.
\newblock In {\em Computer Vision, 2001. ICCV 2001. Proceedings. Eighth IEEE
  International Conference on}, volume~2, pages 416--423. IEEE, 2001.

\bibitem{park2020fast}
Seobin Park, Jinsu Yoo, Donghyeon Cho, Jiwon Kim, and Tae~Hyun Kim.
\newblock Fast adaptation to super-resolution networks via meta-learning.
\newblock In {\em Computer Vision--ECCV 2020: 16th European Conference,
  Glasgow, UK, August 23--28, 2020, Proceedings, Part XXVII 16}, pages
  754--769. Springer, 2020.

\bibitem{Rad_2019_ICCV}
Mohammad~Saeed Rad, Behzad Bozorgtabar, Urs-Viktor Marti, Max Basler,
  Hazim~Kemal Ekenel, and Jean-Philippe Thiran.
\newblock Srobb: Targeted perceptual loss for single image super-resolution.
\newblock In {\em The IEEE International Conference on Computer Vision (ICCV)},
  October 2019.

\bibitem{sajjadi2017enhancenet}
Mehdi~SM Sajjadi, Bernhard Sch{\"o}lkopf, and Michael Hirsch.
\newblock Enhancenet: Single image super-resolution through automated texture
  synthesis.
\newblock In {\em Computer Vision (ICCV), 2017 IEEE International Conference
  on}, pages 4501--4510. IEEE, 2017.

\bibitem{shocher2018zero}
Assaf Shocher, Nadav Cohen, and Michal Irani.
\newblock “zero-shot” super-resolution using deep internal learning.
\newblock In {\em Proceedings of the IEEE conference on computer vision and
  pattern recognition}, pages 3118--3126, 2018.

\bibitem{soh2020meta}
Jae~Woong Soh, Sunwoo Cho, and Nam~Ik Cho.
\newblock Meta-transfer learning for zero-shot super-resolution.
\newblock In {\em Proceedings of the IEEE/CVF Conference on Computer Vision and
  Pattern Recognition}, pages 3516--3525, 2020.

\bibitem{timofte2017ntire}
Radu Timofte, Eirikur Agustsson, Luc Van~Gool, Ming-Hsuan Yang, and Lei Zhang.
\newblock Ntire 2017 challenge on single image super-resolution: Methods and
  results.
\newblock In {\em Proceedings of the IEEE conference on computer vision and
  pattern recognition workshops}, pages 114--125, 2017.

\bibitem{tong2017image}
Tong Tong, Gen Li, Xiejie Liu, and Qinquan Gao.
\newblock Image super-resolution using dense skip connections.
\newblock In {\em Computer Vision (ICCV), 2017 IEEE International Conference
  on}, pages 4809--4817. IEEE, 2017.

\bibitem{wang2021unsupervised}
Longguang Wang, Yingqian Wang, Xiaoyu Dong, Qingyu Xu, Jungang Yang, Wei An,
  and Yulan Guo.
\newblock Unsupervised degradation representation learning for blind
  super-resolution.
\newblock In {\em Proceedings of the IEEE/CVF Conference on Computer Vision and
  Pattern Recognition}, pages 10581--10590, 2021.

\bibitem{wang2021real}
Xintao Wang, Liangbin Xie, Chao Dong, and Ying Shan.
\newblock Real-esrgan: Training real-world blind super-resolution with pure
  synthetic data.
\newblock In {\em Proceedings of the IEEE/CVF International Conference on
  Computer Vision}, pages 1905--1914, 2021.

\bibitem{wang2018recovering}
Xintao Wang, Ke Yu, Chao Dong, and Chen~Change Loy.
\newblock Recovering realistic texture in image super-resolution by deep
  spatial feature transform.
\newblock {\em arXiv preprint arXiv:1804.02815}, 2018.

\bibitem{Wang_2018_ECCV_Workshops}
Xintao Wang, Ke Yu, Shixiang Wu, Jinjin Gu, Yihao Liu, Chao Dong, Yu Qiao, and
  Chen Change~Loy.
\newblock Esrgan: Enhanced super-resolution generative adversarial networks.
\newblock In {\em The European Conference on Computer Vision (ECCV) Workshops},
  September 2018.

\bibitem{wei2020aim}
Pengxu Wei, Hannan Lu, Radu Timofte, Liang Lin, Wangmeng Zuo, Zhihong Pan,
  Baopu Li, Teng Xi, Yanwen Fan, Gang Zhang, et~al.
\newblock Aim 2020 challenge on real image super-resolution: methods and
  results.
\newblock In {\em European Conference on Computer Vision}, pages 392--422.
  Springer, 2020.

\bibitem{wenlong2021ranksrgan}
Zhang Wenlong, Liu Yihao, Chao Dong, and Yu Qiao.
\newblock Ranksrgan: Generative adversarial networks with ranker for image
  super-resolution.
\newblock {\em IEEE Transactions on Pattern Analysis and Machine Intelligence},
  2021.

\bibitem{xie2021finding}
Liangbin Xie, Xintao Wang, Chao Dong, Zhongang Qi, and Ying Shan.
\newblock Finding discriminative filters for specific degradations in blind
  super-resolution.
\newblock {\em arXiv preprint arXiv:2108.01070}, 2021.

\bibitem{yuan2018unsupervised}
Yuan Yuan, Siyuan Liu, Jiawei Zhang, Yongbing Zhang, Chao Dong, and Liang Lin.
\newblock Unsupervised image super-resolution using cycle-in-cycle generative
  adversarial networks.
\newblock In {\em Proceedings of the IEEE Conference on Computer Vision and
  Pattern Recognition Workshops}, pages 701--710, 2018.

\bibitem{zeyde2010single}
Roman Zeyde, Michael Elad, and Matan Protter.
\newblock On single image scale-up using sparse-representations.
\newblock In {\em International conference on curves and surfaces}, pages
  711--730. Springer, 2010.

\bibitem{zhang2012single}
Kaibing Zhang, Xinbo Gao, Dacheng Tao, Xuelong Li, et~al.
\newblock Single image super-resolution with non-local means and steering
  kernel regression.
\newblock {\em Image}, 11:12, 2012.

\bibitem{zhang2021designing}
Kai Zhang, Jingyun Liang, Luc Van~Gool, and Radu Timofte.
\newblock Designing a practical degradation model for deep blind image
  super-resolution.
\newblock {\em arXiv preprint arXiv:2103.14006}, 2021.

\bibitem{zhang2018learning}
Kai Zhang, Wangmeng Zuo, and Lei Zhang.
\newblock Learning a single convolutional super-resolution network for multiple
  degradations.
\newblock In {\em Proceedings of the IEEE Conference on Computer Vision and
  Pattern Recognition}, pages 3262--3271, 2018.

\bibitem{zhang2019ranksrgan}
Wenlong Zhang, Yihao Liu, Chao Dong, and Yu Qiao.
\newblock Ranksrgan: Generative adversarial networks with ranker for image
  super-resolution.
\newblock In {\em Proceedings of the IEEE/CVF International Conference on
  Computer Vision}, pages 3096--3105, 2019.

\bibitem{zhang2018rcan}
Yulun Zhang, Kunpeng Li, Kai Li, Lichen Wang, Bineng Zhong, and Yun Fu.
\newblock Image super-resolution using very deep residual channel attention
  networks.
\newblock In {\em ECCV}, 2018.

\bibitem{zhang2018residual}
Yulun Zhang, Yapeng Tian, Yu Kong, Bineng Zhong, and Yun Fu.
\newblock Residual dense network for image super-resolution.
\newblock In {\em The IEEE Conference on Computer Vision and Pattern
  Recognition (CVPR)}, 2018.

\end{thebibliography}
